\documentstyle[12pt]{article} 
\begin{document}
\begin{center}
{\bf  NON-NOETHER SYMMETRIES IN SINGULAR DYNAMICAL SYSTEMS}\\
\vspace*{5mm}
{\bf George Chavchanidze}

{\scriptsize Department of Theoretical Physics\\ 
A. Razmadze Institute Mathematics\\
1 Aleksidze Street, Ge 380093\\ 
Tbilisi, Georgia\\ e-mail:gch@rmi.acnet.ge}
\end{center}

\begin{abstract}
In the present paper geometric aspects of relationship
between non-Noether symmetries and conservation laws in Hamiltonian
systems is discussed. Case of irregular/constrained dynamical systems
on presymplectic and Poisson manifolds is considered.\\
{\bf 2000 Mathematical Subject Classification:} 70H33, 70H06, 53Z05
\end{abstract}

\section{Introduction}
Noether's theorem associates conservation laws with particular continuous symmetries of
the Lagrangian. According to the Hojman's theorem [1-3] there exists the definite correspondence between
non-Noether symmetries and conserved quantities. In 1998 M. Lutzky showed that several integrals of
motion might correspond to a single one-parameter group of non-Noether transformations
[4]. In
the present paper, the extension of Hojman-Lutzky theorem to singular dynamical systems is considered.
\\
First of all let us recall some basic knowledge of description of the regular dynamical systems
(see, e. g.,[5]).
In this case time evolution is governed by Hamilton's equation
\begin{eqnarray}
i_{X_{h}}\omega + dh = 0,
\end{eqnarray}
where $\omega $ is the closed
($d\omega = 0$) and non-degenerate
($i_{X}\omega = 0 \rightarrow X = 0$) 2-form,
$h$ is the Hamiltonian and
$i_{X}\omega $ denotes contraction of
$X$ with $\omega $.
Since $\omega $ is non-degenerate, this gives rise to an isomorphism between the vector
fields and 1-forms given by $i_{X}\omega + \alpha = 0$.
The vector field is said to be Hamiltonian if it corresponds to exact form
\begin{eqnarray}
i_{X_{f}}\omega + df = 0.
\end{eqnarray}
The Poisson bracket is defined as follows:
\begin{eqnarray}
\{f , g\} = X_{f} g = - X_{g} f = i_{X_{f}}
i_{X_{g}}\omega .
\end{eqnarray}
By introducing a bivector field $W$ satisfying
\begin{eqnarray}
i_{X}i_{Y}\omega = i_{W} i_{X}\omega \wedge i_{Y}\omega ,
\end{eqnarray}
Poisson bracket can be rewritten as
\begin{eqnarray}
\{f , g\} = i_{W} df\wedge dg.
\end{eqnarray}
It's easy to show that
\begin{eqnarray}
i_{X}i_{Y}L_{Z}\omega =
i_{[Z,W]} i_{X}\omega \wedge i_{Y}\omega ,
\end{eqnarray}
where the bracket $[~,~]$ is actually a supercommutator
(for an arbitrary bivector field
$ W = \Sigma _{i} V^{i}\wedge U^{i} $ we have
$[X,W] = \Sigma _{i}[X,V^{i}]\wedge U^{i}
+ \Sigma _{i}V^{i}\wedge [X,U^{i}] $).
Equation (6) is based on the following useful property of the Lie derivative
\begin{eqnarray}
L_{X}i_{W}\omega = i_{[X,W]}\omega +
i_{W}L_{X}\omega .
\end{eqnarray}
Indeed, for an arbitrary bivector field
$W = \Sigma _{i} V^{i}\wedge U^{i} $ we have
\begin{eqnarray}
L_{X}i_{W}\omega =
L_{X}i_{\Sigma _{i}
V^{i}\wedge U^{i}}\omega =
L_{X}\Sigma _{i}
i_{U^{i}}i_{V^{i}}\omega =
 \nonumber \\
= \Sigma _{i}
i_{[X,U^{i}]}i_{V^{i}}\omega +
\Sigma _{i}
i_{U^{i}}i_{[X,V^{i}]}\omega +
\Sigma _{i}
i_{U^{i}}i_{V^{i}}L_{X}\omega =
i_{[X,W]}\omega + i_{W}L_{X}\omega
\end{eqnarray}
where $L_{Z}$ denotes the Lie derivative along the vector field $Z$.
According to Liouville's theorem Hamiltonian vector field
preserves $\omega $
\begin{eqnarray}
L_{X_{f}}\omega = 0;
\end{eqnarray}
therefore it commutes with $W$:
\begin{eqnarray}
[X_{f} ,W] = 0.
\end{eqnarray}
In the local coordinates $ z_{i} $ where
$\omega = \omega ^{ij}dz_{i}\wedge z_{j}$ bivector field
$W$ has the following form
$W = W^{ij}\partial _{z_{i}}\wedge \partial _{z_{j}}$ where
$W^{ij}$ is matrix inverted to $\omega ^{ij}$.
\\
\section{ Case of regular Lagrangian systems}
We can say that a group of transformations
$g(a) = e^{aL_{E}}$ generated by the vector
field $E$ maps the space of solutions of equation onto itself if
\begin{eqnarray}
i_{X_{h}}g_{*}(\omega )
+ g_{*}(dh) = 0
\end{eqnarray}
For $X_{h}$ satisfying
\begin{eqnarray}
i_{X_{h}}\omega + dh = 0
\end{eqnarray}
Hamilton's equation.
It's easy to show that the vector field $E$ should satisfy
$[E , X_{h}] = 0$
(Indeed,
$i_{X_{h}}L_{E}\omega + dL_{E}h =
L_{E}(i_{X_{h}}\omega + dh) = 0$
since $[E,X_{h}] = 0$). When $E$ is not Hamiltonian,
the group of transformations $g(a) = e^{aL_{E}}$ is non-Noether
symmetry (in a sense that it maps solutions onto solutions but does not preserve action).
\\
{\bf Theorem:} (Lutzky, 1998) If the vector field $E$ generates non-Noether symmetry, 
then the following functions are constant along solutions:
\begin{eqnarray}
I^{(k)} = i_{W^{k}} \omega _{E}^{k} ~~~~ k = 1...n,
\end{eqnarray}
where $W^{k}$ and $\omega _{E}^{k}$ are outer
powers of $W$ and $L_{E}\omega $.
\\
{\bf Proof:} We have to prove that $I^{(k)}$ is constant along
the flow generated by the Hamiltonian. In other words, we should find that
$L_{X_{h}}I^{(k)} = 0$ is
fulfilled. Let us consider
$L_{X_{h}}I^{(1)}$
\begin{eqnarray}
L_{X_{h}}I^{(1)}
= L_{X_{h}}(i_{W}\omega _{E}) =
i_{[X_{h} , W]}\omega _{E}
+ i_{W}L_{X_{h}}\omega _{E},
\end{eqnarray}
where according to Liouville's theorem both terms
($[X_{h} , W] = 0$ and
$i_{W}L_{X_{h}}L_{E}\omega =
i_{W}L_{E}L_{X_{h}}\omega =
0$ since $[E , X_{h}] =
0$ and $L_{X_{h}}\omega = 0$) vanish.
In the same manner one can verify that
$L_{X_{h}}I^{(k)} = 0~~ $
\\
{\bf Note 1:} Theorem is valid for a larger class of generators $E$ .
Namely, if $[E , X_{h}] = X_{f}$ where $X_{f}$ is
an arbitrary Hamiltonian vector field, then $I^{(k)}$ is still conserved. Such a
symmetries map the solutions of the equation
$i_{X_{h}}\omega + dh = 0$
on solutions of
$i_{X_{h}}g_{*}(\omega ) +
d(g_{*}h + f) = 0$.
\\
{\bf Note 2:} Discrete non-Noether symmetries give rise to the conservation of
$I^{(k)} = i_{W^{k}}g_{*}(\omega )^{k}$
where $g_{*}(\omega )$ is transformed $\omega $.
\\
{\bf Note 3:} If $I^{(k)}$ is a set of conserved quantities
associated with $E$ and $f$ is any conserved quantity, then the set of functions
$\{I^{(k)} , f\} $
(which due to the Poisson theorem are integrals of motion) is associated with
$[X_{h} , E]$. Namely it is easy to show by taking the Lie
derivative of (13) along vector field $ E$ that
$\{I^{(k)} , f\} =
i_{W^{k}}\omega ^{k}_{[X_f , E]}$ is fulfilled.
As a result conserved quantities associated with Non-Noether symmetries form Lie algebra under
the Poisson bracket.
\\
{\bf Note 4:}
If generator of symmetry satisfies Yang-Baxter equation
$[[E[E , W]]W] = 0$ Lutzky's conservation laws are in involution [7]
$\{Y^{(l)} , Y^{(k)}\} = 0$\\
\section{ Case of irregular Lagrangian systems}
The singular Lagrangian (Lagrangian with vanishing Hessian) leads to degenerate 2-form
$\omega $ and we no longer have isomorphism between vector fields and 1-forms.
Since there exists a set of "null vectors" $u^{k}$ such that
$i_{u^{k}}\omega = 0~ ~k = 1,2 ... n - rank(\omega ),$
every Hamiltonian vector field is
defined up to linear combination of vectors $u^{k}$. By identifying $X_{f}$
with $X_{f} + C_{k}u^{k},$ we can introduce equivalence class
$ X_{f}^{\bullet }$ (then all $u^{k}$ belong to
$0^{\bullet }$ ).
The bivector field $W$ is also far from being unique, but if
$W_{1}$ and $W_{2}$ both satisfy
\begin{eqnarray}
i_{X}i_{Y} \omega =
i_{W_{1,2}} i_{X}\omega \wedge i_{Y}\omega ,
\end{eqnarray}
then
\begin{eqnarray}
i_{(W_{1}
- W_{2})} i_{X}\omega \wedge i_{Y}\omega =
0 ~~~\forall X,Y
\end{eqnarray}
is fulfilled. It is possible only when
\begin{eqnarray}
W_{1} - W _{2} = v_{k}\wedge u^{k}
\end{eqnarray}
where $v_{k}$ are some vector fields and
$i_{u^{k}}\omega = 0$
(in other words when $ W_{1} - W_{2}$ belongs to the class
$0^{\bullet }$)
\\
{\bf Theorem:} If the non-Hamiltonian vector field $E$
satisfies $[E , X_{h}^{\bullet }] = 0^{\bullet } $ commutation
relation (generates non-Noether symmetry), then the functions
\begin{eqnarray}
I ^{(k)}
= i_{W^{k}}\omega _{E}^{k} ~~~~ k = 1...rank(\omega )
\end{eqnarray}
(where $\omega _{ E} = L_{E}\omega $) are constant along trajectories.
\\
{\bf Proof:} Let's consider $ I^{(1)}$
\begin{eqnarray}
L_{X_{h}^{\bullet }}I^{(1)}
= L_{X_{h}^{\bullet }}(i_{W}\omega _{E})
= i_{[X_{h}^{\bullet } , W]}\omega _{E} +
i_{W}L_{X_{h}^{\bullet }}\omega _{E} = 0
\end{eqnarray}
The second term vanishes since $[E , X_{h}^{\bullet }] = 0^{\bullet }$ and
$L_{X_{h}^{\bullet }}\omega = 0$. The first one is
zero as far as $ [X_{h}^{\bullet } , W^{\bullet }] = 0^{\bullet }$ and
$[E , 0^{\bullet }] = 0^{\bullet }$ are satisfied. So
$I^{ (1)}$ is conserved.
Similarly one can show that $L_{X_{h}}I^{(k)} = 0$ is
fulfilled.
\\
{\bf Note 1:} $W$ is not unique, but $I^{(k)}$ doesn't depend
on choosing representative from the class $W^{\bullet }$.
\\
{\bf Note 2:} Theorem is also valid for generators $E$ satisfying
$ [E , X_{h}^{\bullet }] = X_{f}^{\bullet }$
\\
{\bf Example:} Hamiltonian description of the relativistic particle leads to the following action
\begin{eqnarray}
A = \int (p^{2} + m^{2})^{\frac{1}{2}}dx_{0}
+ p_{k}dx_{k}
\end{eqnarray}
with vanishing canonical Hamiltonian and degenerate 2-form
\begin{eqnarray}
\omega = (p^{2} +
m^{2})^{-\frac{1}{2}} (p_{k}dp_{k}\wedge dx_{0}
+ (p^{2} + m^{2})^{\frac{1}{2}}dp_{k}\wedge dx_{k}).
\end{eqnarray}
$\omega $ possesses the "null vector field"
$i_{u}\omega = 0$
\begin{eqnarray}
u = (p^{2} + m^{2})^{\frac{1}{2}}\partial _{x_{0}}
+ p_{k}\partial _{x_{k}}.
\end{eqnarray}
One can check that the following non- Hamiltonian vector field
\begin{eqnarray}
E = (p^{2} + m^{2})^{\frac{1}{2}}x_{0}\partial _{x_{0}}
+ p_{1}x_{1}\partial _{x_{1}} + ... + p_{k}x_{n}\partial _{x_{n}}
\end{eqnarray}
generates non-Noether symmetry. Indeed, $E$ satisfies
$[E , X_{h}^{\bullet }] = 0^{\bullet }$ because of
$X_{h}^{\bullet } = 0^{\bullet }$ and $[E,u] = u$.
Corresponding integrals of motion are combinations of momenta:
\begin{eqnarray}
I^{(1)} = (p^{2} + m^{2})^{\frac{1}{2}} + p_{1} + ...
+ p_{k} = \sum _{a}p_{a};
\\
I^{(2)} = \sum _{a\neq b} p_{a}p_{b};
\\
...
\\
I^{(n)} = \prod _{a}p_{a}
\end{eqnarray}
This example shows that the set of conserved quantities can be obtained from a single
one-parameter group of non-Noether transformations.
\\
\section{Case of dynamical systems on Poisson Manifold}
The previous two sections dealt with dynamical systems on symplectic and presymplectic manifolds.
Now let us consider the case of dynamical systems on the Poisson manifold. In general, the Poisson
manifold is an even dimensional manifold equipped with the Poisson bracket which can be defined by
means of the bivector field $W$ satisfying $[W,W] = 0$ as follows:
\begin{eqnarray}
\{f , g\} = i_{W} df\wedge dg
\end{eqnarray}
Due to skewsymmetry of the $W$ Poisson bracket is also skewsymmetric and, in general, it is
degenerate. The commutation relation $[W,W] = 0$ (where
$[~,~]$ denotes the supercommutator of vector fields) ensures that
the Poisson bracket satisfies the Jacobi identity.
As well as in case of symplectic (presymplectic) manifold, we have
correspondence between vector fields and 1-forms governed by equation:
\begin{eqnarray}
\beta (X) + \alpha \wedge \beta (W) = 0
~~~\forall \beta 
\end{eqnarray}
The classification of vector fields is based on this correspondence.
The vector field is called the (locally) Hamiltonian if it corresponds to the (closed) exact 1-form
\begin{eqnarray}
\beta (X_{h}) + dh\wedge \beta (W) = 0
~~~\forall \beta 
\end{eqnarray}
and the non-Hamiltonian if the corresponding 1-form is not closed (Note that, when
$W$ is degenerate there are vector fields that are not assosiated with 1-forms and lay
beyond our classification).
Note also that, according to the Liouville's theorem, Hamiltonian vector field
preserves the bivector field $W$, i. e. ,
$L_{X_{h}}W = [X_{h},W] = 0$.
\\
Now let's consider one-parameter group of transformations
$g(a) = e^{aL_{E}}$
generated by the non-Hamiltonian vector field $E$.
Like in the regular case, $E$
generates symmetry of Hamilton's equation (maps space of solutions onto itself) if
$[E,X_{h}] = 0$ and the correspondence between non-Noether
symmetries and conservation laws is governed by the following theorem:
\\
{\bf Theorem:} If the non-Hamiltonian vector field $E$ generates the symmetry
of Hamilton's equation, then the set of functions
\begin{eqnarray}
I^{(k)} = \frac{[E , W]^{r - k}\wedge W^{k}}{W^{r}}
\end{eqnarray}
(where $r$ is the rank of the bivector field $W$ and $k = 1,2, ... r$)
is conserved.
\\
{\bf Proof:} It is clear that $I^{k}$ are conserved, since
$W$ and $[E , W]$ are invariant bivector fields
($L_{X_{h}} = [X_{h},W] = 0$, according to the
Liouville's theorem and
\begin{eqnarray}
L_{X_{h}}[E , W] =
[X_{h}[E , W]] = [[X_{h}, E]W] =
[E[X_{h} , W]] = 0,
\end{eqnarray}
since $[E,X_{h}]
= 0$)
Now we have to show that the ratio of r-vector fields
\begin{eqnarray}
[E , W]^{r - k}\wedge W^{k} k = 0, 1, ... r,
\end{eqnarray}
is defined correctly. Explicitly this fact can be demonstrated in (local) cannonical
coordinates where
$W =
\sum _{i = 1}^{r}\partial _{p_{i}}\wedge \partial _{q_{i}}$
and every non-Hamiltonian vector field can be represented as
$E =
\sum _{i = 1}^{r} E^{(p_{i})}\partial _{p_{i}} +
E^{(q_{i})}\partial _{q_{i}}$ and as a result every r-vector field
of the form (33) is proportional to
$w = 
\partial _{p_{1}}\wedge \partial _{q_{1}}\wedge \partial _{p_{2}}\wedge \partial _{q_{2}}\wedge ...
\partial _{p_{r}}\wedge \partial _{q_{r}}$
{\bf q. e. d.}
\\
\section{Acknowledgements}
{\it Author is grateful to Z. Giunashvili and M. Maziashvili for
constructive discussions and particularly grateful to George Jorjadze for invaluable help.
This work was supported by INTAS (00-00561)
and Scholarship from World Federation of Scientists.}
\\

\end{document}